\documentclass[11pt,a4paper]{article}
\usepackage{txfonts}
\usepackage{cospar}

\usepackage[sectionbib]{natbib}

\usepackage{url}
\usepackage{graphicx}
\def \rns{R_{\scriptscriptstyle NS}}
\def \delp {\Delta^{\rm peak}}

\title{SIGNATURES OF PULSAR POLAR-CAP EMISSION AT THE HIGH-ENERGY SPECTRAL 
CUT-OFF}

\author{J. Dyks\address{Laboratory for High Energy Astrophysics, NASA/GSFC, 
        Greenbelt, MD 20771,
USA}$^{,2}$
        and B. Rudak\address{Nicolaus Copernicus Astronomical Center, 
        Rabia{\' n}ska 8, 87-100 Toru{\' n}, Poland}$^,$\address{Dept. of Astronomy and Astrophysics, NCU, Toru{\' n}, Poland}}

\begin{document}

\maketitle

\begin{abstract}
We address four unique signatures in pulsar gamma-ray radiation
as predicted by polar-cap models. 
These signatures are expected to be present nearby the spectral high-energy cutoff at around several GeV.
Their magnitude  and, therefore, their observability depends strongly
on the orientational factors, the rotation,  as well as on the details of 
the polar cap structure.
These strong predictions are likely to be verified by the NASA's future gamma-ray mission GLAST.
\end{abstract}

\section*{INTRODUCTION}
The properties of high-energy radiation from pulsars challenge all theoretical models
of pulsar magnetospheres and stimulate their development (for recent reviews on the polar cap scenario see
Rudak et al., 2002 and Baring, 2003).
The aim of this contribution is to address four characteristic features 
in the gamma-ray radiation induced by rotation 
as well as by orientational (viewing and inclination) factors. These features are 
thought to be unique signatures of 
the polar-cap activity in gamma-ray pulsars.

Due to the limited space we concentrate in this contribution on the issue of the spectral shape at the high-energy cutoff.
The other three signatures are discussed in a lesser extent, but the references are given to their elaborate treatment. 

\begin{figure}[!t]
\vskip -0.25cm
\includegraphics[width=0.9\textwidth]{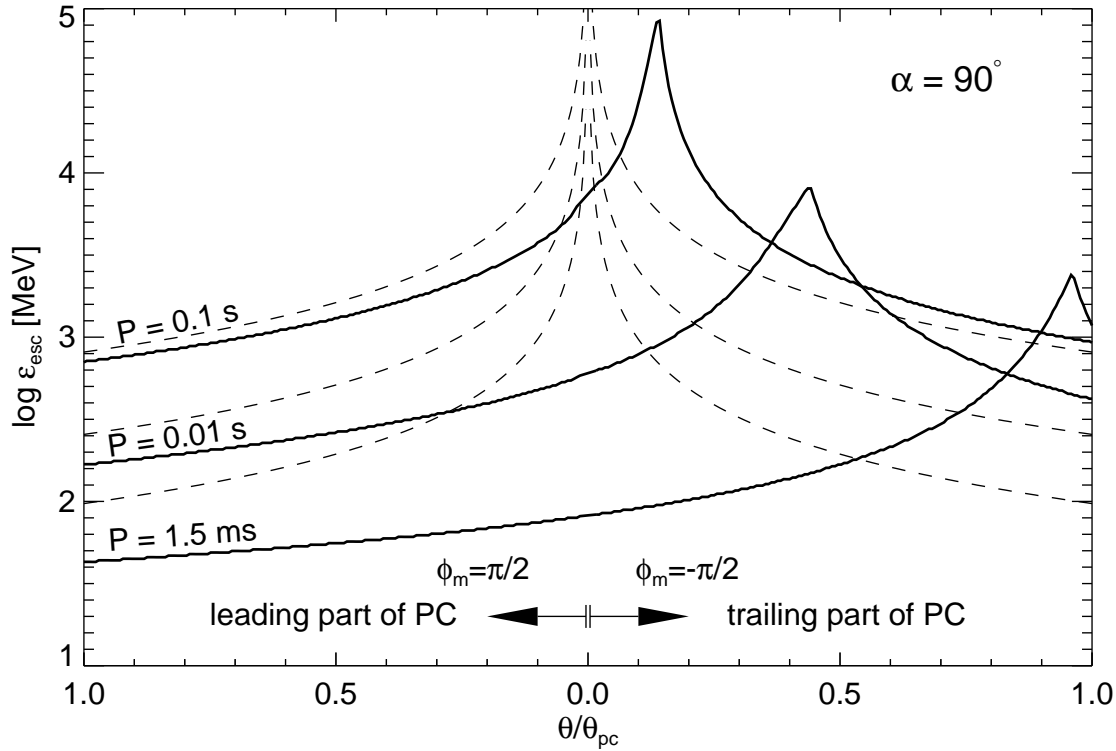}
\vskip -1.cm
\caption{
The escape energy $\varepsilon_{\rm esc}$ of photons from the polar cap
surface of an orthogonal rotator
with $B_{\rm pc} = 10^{12}$G
is shown as a function of normalized magnetic colatitude $\theta/\theta_{\rm
pc}$ of the emission points.
The points are assumed to lay along the cross-section of the polar cap
surface
with the equatorial plane of rotation, thus location of each point is
determined by $\theta/\theta_{\rm pc}$
in the range $[0,\,1]$, and the magnetic azimuth $\phi_{\rm m}$  equal
either to
$\pi/2$ (for the leading half of the polar cap) or $-\pi/2$ (for the
trailing half).
Three solid lines
are labelled with the corresponding spin periods $P$ of $0.1$~s, $10$~ms,
and $1.5$~ms.
Each solid line is accompanied by a dashed line 
calculated for the case when rotational effects are ignored.
}
\label{f1}
\end{figure}

\section*{SUPER-EXPONENTIAL SHAPE OF THE HIGH-ENERGY CUTOFF}
It is generally accepted that an inherent feature in polar cap models is a very sharp,
super-exponential cutoff at the
high-energy  end of the pulsar spectrum (e.g. de Jager, 2002).
Thompson (2001) argues that GLAST will be able to pinpoint such a signature
in phase-averaged spectra. One would then have (hopefully) a direct evidence of a one-photon
magnetic absorption taking place deep in inner parts of the magnetosphere;
alternatively - the cutoff with an exponential shape would speak for outer-gap
activity.
However, the latter interpretation  may well be premature for two reasons:
First, the shape of high-energy cutoff in pulsar spectra does
depend on viewing geometry (Harding and Zhang, 2001). In  off-beam cases (i.e. when the line of sight
misses the highest energy gamma-ray beam) the cutoff has a simple
exponential shape due to the upper limit in the energy distribution of those particles which emit
observable photons
(Rudak et al., 2002). 
\begin{figure}[!t]
\hskip 0.6cm
\includegraphics[width=0.92\textwidth]{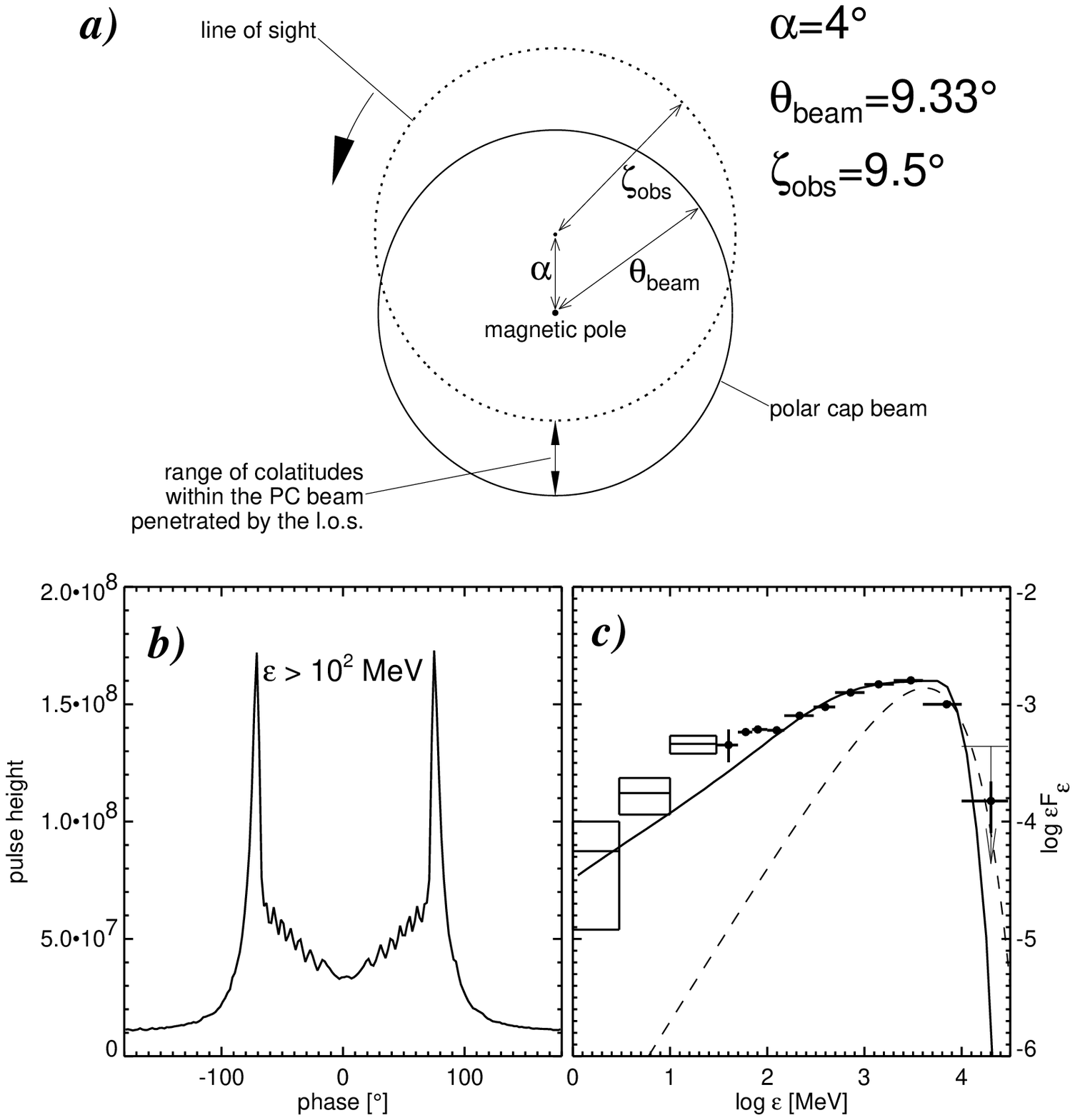}
\vskip -1.5cm
\caption{
a) Viewing geometry assumed to calculate the gamma-ray lightcurve and the
spectrum shown in panels b) and c). The line of sight
penetrates the polar cap beam  (within the solid-line circle) along the dotted-line trajectory.
b) Lightcurve calculated for $\epsilon > 10^2$ MeV, with a peak separation of $0.42$ in phase (the model refers to
the Vela pulsar). 
c) Phase-averaged spectrum
(solid line) of the model, with the COMPTEL and EGRET data points for Vela also indicated. 
Instantaneous spectrum of curvature radiation with exponential-like cutoff due to monoenergetic particles is shown
for comparison (dashed line). 
Above $\sim 10$~GeV the modelled spectrum assumes a super-exponential cutoff, which is  much sharper than the simple exponential. 
}
\label{f2}
\end{figure}
Fortunately, the off-beam cases should
be easily identified by virtue of their spectral softness relative to the on-beam pulsars
- their energy spectra peak typically around $100$~MeV 
hardly extending beyond $1$~GeV (see Figures 2 to 5 in Wo{\' z}na et al., 2002).
Second, even in the case of on-beam geometry (i.e. when the line of sight penetrates the polar cap)  the
apparent cutoff in phase-averaged spectrum may well assume a quasi-exponential shape. 
This is possible for the so called filled-column beam cases (i.e. when
emission from the interior of the polar cap is non-negligible).
To show this possibility we present two almost identical filled-column beam models, 
with just two different values for the inclination angle $\alpha$ between the spin and the magnetic axis
(see Figures \ref{f2} and \ref{f3}). 
We assumed the source of primary electrons
to be uniform over the polar cap,
with instant acceleration.
For the radiation emitted at different distances from magnetic axis
(magnetic colatitudes $\theta$)
the sharp cutoff occurs at different photon energies often denoted as escape energies $\varepsilon_{\rm esc}$.
The escape energy $\varepsilon_{\rm esc}$ strongly depends on the magnetic colatitude as shown in Figure \ref{f1}
(see for details Dyks and Rudak, 2002).
In the course of pulsar rotation the line of sight samples various magnetic 
colatitudes,
and therefore, the phase-averaged spectrum is composed of many spectra, each
with a different position of the cutoff.
[Note, that the depth of the interpulse dip at phase $0^{\rm o}$ of the pulse profile is much more pronounced 
in Figure 3b than in Figure 2b.
This is because in the former case the line of sight 
penetrates the regions of smaller magnetic colatitudes - with
larger curvature radii of the magnetic field lines
and, therefore, with lower level of curvature emission.]
When the range of sampled colatitudes is narrow (cf. Figure \ref{f2}) the phase averaged
spectrum does have a much sharper cutoff than a simple
exponential decay. However, for a broader range of sampled
colatitudes (cf. Figure \ref{f3}), 
the cutoff in the phase-averaged spectrum may look 
exactly like a simple exponential decay.
\begin{figure}[!t]
\hskip 0.6cm
\includegraphics[width=0.92\textwidth]{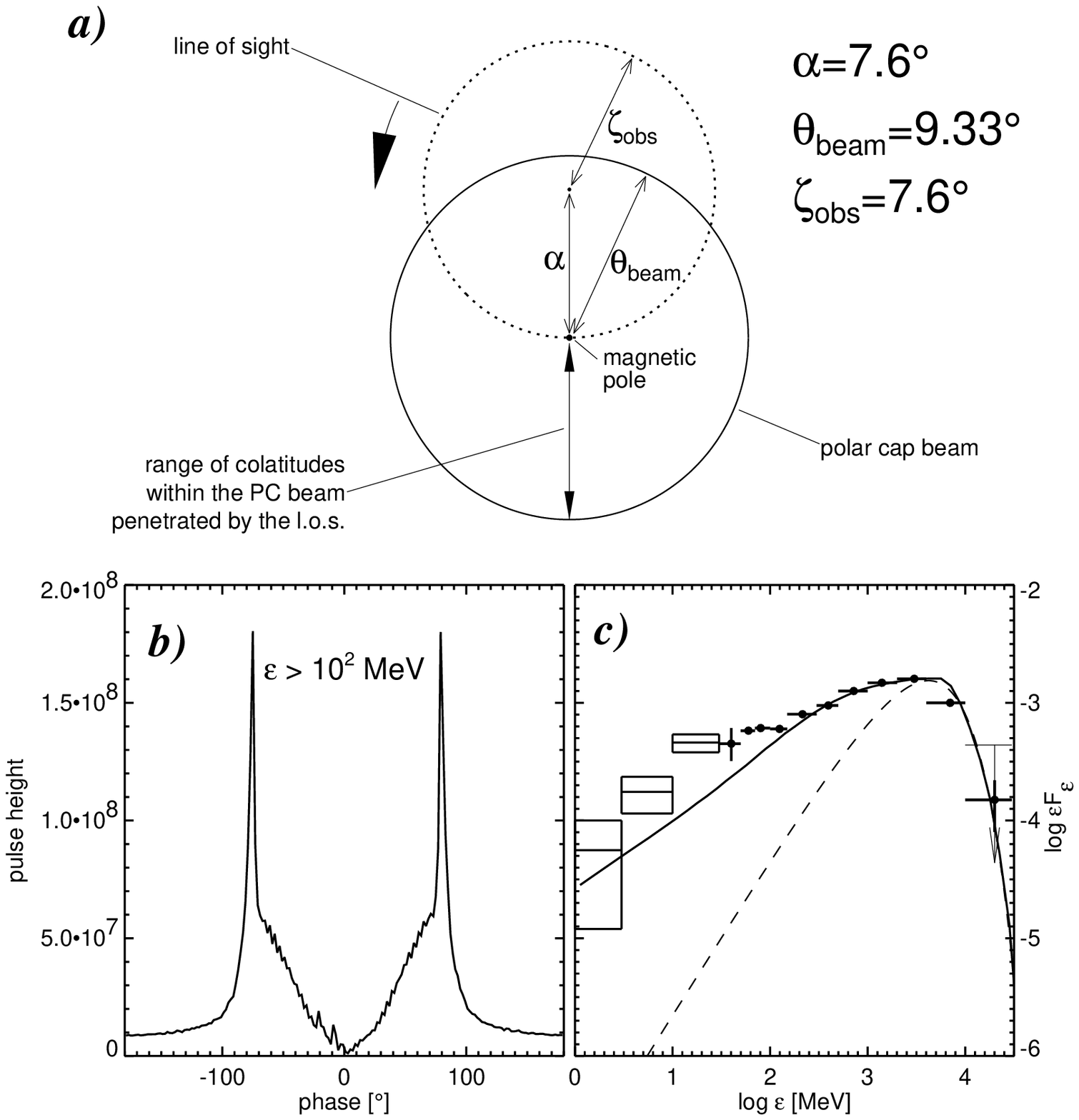}
\vskip -1.5cm
\caption{
Same as in Figure \ref{f2} but for a different orientational set-up: the inclination angle $\alpha$
and the viewing angle $\zeta$ (the angle between the spin axis and the l.o.s) have changed (but in such a way that the peak separation
in the lightcurve remains unchanged). The line of sight
penetrates now, unlike in the previous case, the innermost parts of the polar cap beam. 
In consequence, the modelled phase-averaged spectrum 
reveals now a simple exponential shape, in contrast to 
the super-exponential shape in
Figure \ref{f2}.
}
\label{f3}
\end{figure}
We conclude that 
within the filled-column beam cases
the apparent shape of the high-energy cutoff in the phase-averaged
spectra depends on the viewing geometry and 
it may emulate simple exponential shape
even for on-beam geometry. To observe sharp super-exponential cutoffs in such cases
one would have to turn to  
phase-resolved spectra.
Alternatively, within the hollow-cone beam cases with emission extended over the large range of altitudes 
(like the slot gap model presented by Muslimov and Harding, 2003) cutoffs in the phase-averaged spectra 
are expected to be much sharper than simple exponential.

\section*{ROTATION-INDUCED ASYMMETRY IN THE DOUBLE-PEAK LIGHTCURVES}
According to the polar cap model, the characteristic double-peak gamma-ray
lightcurve forms when the line of sight intersects the
polar-cap beam where the highest-energy emission originates:
upon entering the beam
a leading peak (LP) is produced, followed by 
a bridge emission due to the inner parts of the beam; when the line of sight exits the beam
the trailing peak (TP) forms.
\begin{figure}[!t]
\hskip 1.0cm
\includegraphics[width=0.85\textwidth]{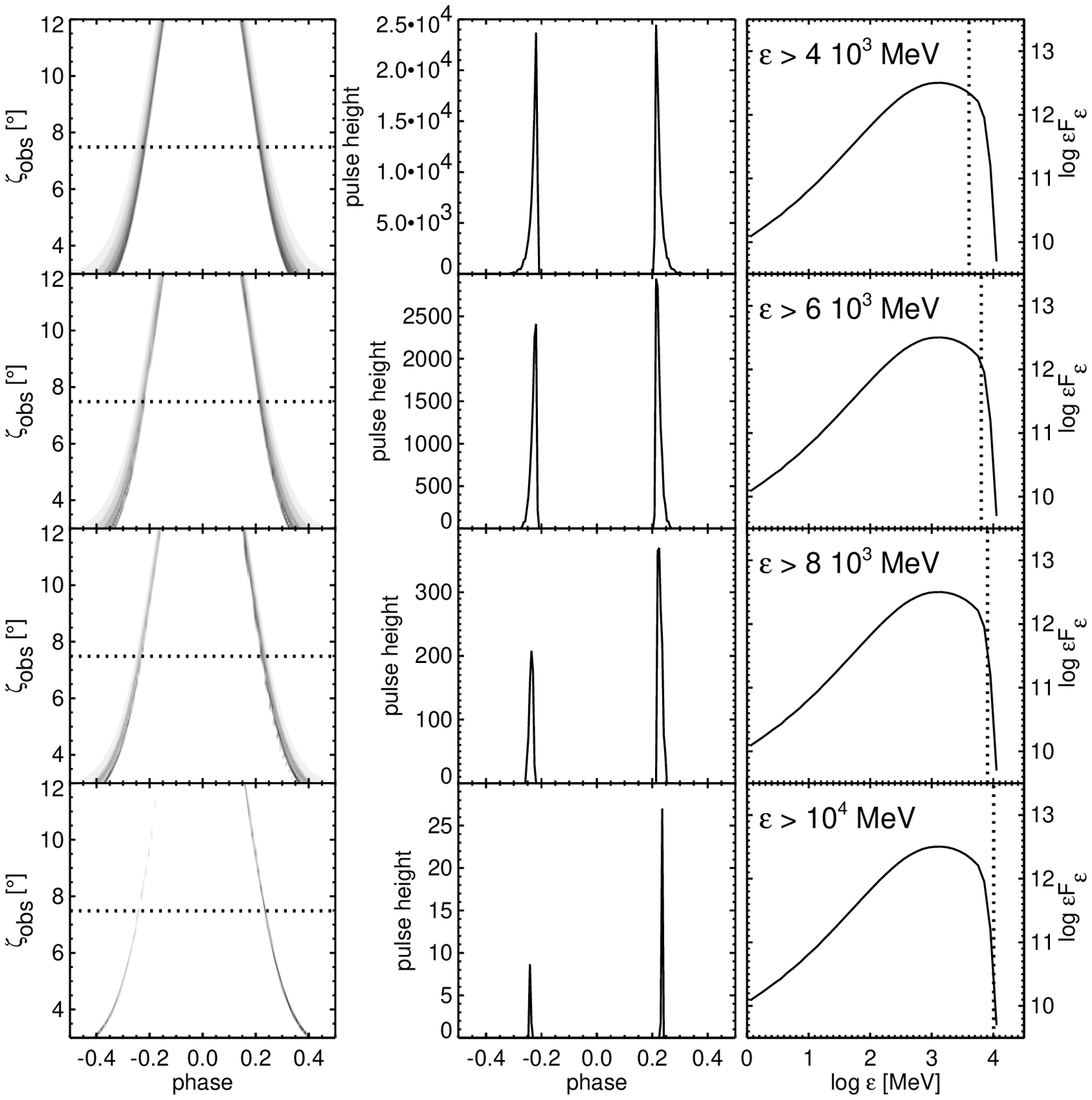}
\vskip -1.cm
\caption{
Gamma-ray radiation characteristics 
calculated for the Vela pulsar with the orientational set-up as shown in panel a) of Figure \ref{f3} but for a hollow-cone beam case
(i.e. with electrons ejected  from the outer rim of the polar
cap).
Left column shows the outgoing photons of energy $\varepsilon > \varepsilon_{\rm limit}$ 
mapped onto the parameter space $\zeta_{\rm obs}$ vs. phase of rotation.
Middle column shows the double-peak lightcurve 
for $\zeta_{\rm obs}=  7.6^\circ$ (yielding the peak separation equal 0.42 in phase).
Right column shows the phase-averaged 
energy spectrum (the y-axis has arbitrary units) for the same
$\zeta_{\rm obs} = 7.6^\circ$. 
Dotted vertical line indicates
the part of the spectrum  ($\varepsilon > \varepsilon_{\rm limit}$) which contributes to the 
corresponding pulse profile in the middle column.
The values of 
$\varepsilon_{\rm limit}$: $4\times 10^3$, $6\times10^3$,
$8\times 10^3$, and $10^4$ MeV, are given in the right column panels.
}
\label{f4}
\end{figure}
It was shown recently (Dyks and Rudak, 2002) that the rotation 
results in an asymmetric magnetic absorption rate  for the leading and the
trailing part of the magnetosphere
even in the case when the magnetic
field structure and the population of radiating particles  are symmetric
around the magnetic pole.
Rotation of the magnetosphere enhances the magnetic absorption of photons
in the leading peak and weakens the absorption of the trailing peak.
Therefore, the leading peak in the lightcurve disappears at a lower photon
energy than the trailing peak, the effect noticed among the brightest EGRET
pulsars by Thompson (2001). 
Figure \ref{f4} shows this effect for the model of the Vela pulsar.

\section*{ROTATION-INDUCED STEP-LIKE SPECTRA}
For the nearly-aligned model of Vela, the weaking of the
leading peak (cf. Figure \ref{f4}) can be discerned only when the acceleration of electrons takes place
at high altitudes, where local corotation velocity is large 
(we took $h = 4\rns$; this seems to be justified by the results of recent work of Muslimov and Harding, 2003). 
For millisecond pulsars, however, with high inclination angles $\alpha$ of the magnetic 
dipole, the difference between cutoff's energy for the leading 
and for the trailing peak becomes huge, and the efect may be noticeable even in
the phase-averaged spectrum as a step-like feature
close to the high-energy cutoff (see Figure 10 of Dyks and Rudak, 2002). Below the step the spectrum consists of photons from
both the leading and the trailing peak, whereas above the step the only photons
which contribute to the spectrum come from the trailing peak. At the step, the level of the spectrum drops by a
factor of two.

\section*{CHANGE OF THE PEAK SEPARATION}
Another interesting consequence of the magnetic absorption of high energy photons is a
noticeable change in the separation $\delp$ between the two peaks in the
double-peak gamma-ray lightcurve when the photon energy approaches the high-energy cutoff
(see Figure 7 of Dyks and Rudak, 2000). 
In models with electrons ejected  from a rim of the polar
cap (hollow cone beam models), 
the higher energy of escaping photons requires
higher emission altitudes to avoid magnetic absorption. 
Because of the nearly-aligned-case orientation,
a slightly larger opening angle of the gamma-ray beam 
translates into a pronounced
increase in the peak separation $\delp$.
If the emission from the interior of the polar cap is included (filled column beam models), the
opposite happens: $\delp$ decreases near the high-energy cutoff in the
spectrum (model D in Figure 7 of Dyks and
Rudak, 2000). This is because photons of the highest energy 
which manage to escape unattenuated  
come in this case from the inner parts of the beam, i.e. closer
to the magnetic dipole axis.
The change of the peak separation is thus especially useful
observational diagnostics: not only does it indicate the polar cap origin 
of the radiation, but also it 
discriminates between the hollow cone and the filled column beam cases.

\section*{CONCLUSIONS} 
The high-energy cutoff range in the pulsar spectrum (above $\sim 5$~GeV)
is potentially a very promising testing ground for the scenario of polar-cap origin of pulsar activity.
There are at least four signatures of the polar-cap activity expected within this range for on-beam pulsars:
1) super-exponential shape of the spectral cutoff (for phase-resolved spectra);
2) weakening of the leading peak in the double peak lightcurves; 
3) abrupt change in the peak separation in the double peak lightcurves;
4) step-like phase-averaged spectra 
for highly inclined millisecond pulsars.
On the other hand, so called off-beam pulsars should
be easily identified by virtue of their spectral softness (with the spectra peaking typically around $\sim 100$~MeV).
The magnitude and the chance for detecting the properties 1) to 4) depends on such
factors as the viewing geometry, 
the details of the polar cap structure and the rotation.
Nevertheless, 
these properties are powerful observational diagnostics
in the context of superior-quality data in the HE and VHE domains
anticipated in the near future from GLAST and MAGIC; one may expect that
at least some of these signatures will be found for a significant number of objects (alternatively: their existence will be
rejected) making thus a strong case for 
the polar-cap scenario of pulsar magnetospheric activity (alternatively: questioning it).

\section*{ACKNOWLEDGEMENTS}
We acknowledge financial support from KBN grant PBZ-KBN-054/P03/2001.

\noindent
e-mail address: bronek@ncac.torun.pl

\end{document}